\documentclass[hyper]{JHEP} 

\usepackage{epsfig}

\newcommand\fverb{\setbox\pippobox=\hbox\bgroup\verb}
\newcommand\fverbdo{\egroup\medskip\noindent%
			\fbox{\unhbox\pippobox}\ }
\newcommand\fverbit{\egroup\item[\fbox{\unhbox\pippobox}]}
\newbox\pippobox
\title{ D-Branes from
$N$ D(-1)-Branes in Bosonic and
Type IIA String Theory.}

\author{by J. Kluso\v{n}\\
	 Department of Theoretical Physics and Astrophysics\\
                   Faculty of Science, Masaryk University\\
Kotl\'{a}\v{r}sk\'{a} 2, 611 37, Brno\\
Czech Republic\\
	E-mail: \email{klu@physics.muni.cz}}

\preprint{\hepth{0102063}}	

\abstract{In this paper we would like to discuss the
emergence of D-branes from  infinite many
D-instantons in bosonic and type IIA string theory
in the framework of boundary string field theory.}

\keywords{D-branes, Matrix models}

\def\tr{\mathrm{Tr}}

\def\bra #1{\left<#1\right|}
\def\ket #1{\left|#1\right>}
\begin{document}

\section{Introduction}
The boundary String Field Theory (BSFT) 
\cite{WittenBT,WittenBT1,WittenBT2,ShatasviliBT,
ShatasviliBT1}
\footnote{In this paper we use the name of the theory
proposed in \cite{KrausBST}.} is a version of open
string field theory in which the classical
configuration space is a space of two-dimensional
world-sheet theories on the disk which are conformal
in the interior of the disk and which have
arbitrary boundary actions. The boundary conformal
theories correspond to the solutions of
classical equations of motion. 
In the recent papers 
\cite{ShatasviliBT2, MooreBT, 
SenBT, Cornalba, Okuyama, MooreBT2,DasguptaBT1,
Aleksejev,Yasnov,Moriyama,
ShatasviliBT3,Kleban,KrausBST,TakayBT} (For the world-sheet
approach to this problem, see \cite{TseytlinSM,HarveyWS1,
HarveyWS2})
 the BSFT was used in the analysis
of the tachyon condensation on the unstable D-brane
systems
in string theories (For review and extensive list of 
references, see \cite{SenR}). It was shown that BSFT is
very effective tool for the study of this problem. In particular,
it was shown that:
\begin{itemize}
\item The condensation to the closed string
vacuum and to lower dimensional branes involves excitations
of only one mode of string field-tachyon.
\item The exact tachyon potential can be computed in BSFT and
its qualitative features agree with the Sen conjecture
\cite{SenC}.
\item The exact tachyon profiles corresponding to the tachyon
condensation to the lower dimensional D-brane give rise to
descent relations between the tensions of various branes 
\cite{Moriyama} which again agree with those expected
\cite{SenC}. 
\end{itemize}
The problem of tachyon condensation was also studied from
the point of view of Witten's open string field theory 
\cite{WittenSFT,SenFT1,SenFT2,BerkovitzFT1,HarveyFT,SenFT3,
TaylorFT,KochFT,Desmet,NaqviFT,SenFT4,DavidFT,WittenFT,
RasteliFT,SenFT5,TaylorFT2,KochFT2,NaqviFT2,Schnabl,Rastelli}. In contrast
with BSFT, the tachyon condensation in general involves 
 giving the expectation values of infinite number of component fields.
As a consequence, only approximate results are available. 

In this note we would like to study the problem of the tachyon
condensation on the system of $N$ D-instantons in the limit 
$N\rightarrow \infty$ in the framework
of BSFT. This system has been studied  in
the low energy approximation in \cite{Kluson1} and
in the language of the Witten's open string field theory in \cite{Kluson2}. 
The problem of the tachyon condensation on the system of
D0-branes has been also studied in \cite{Terashima} from the 
point of view of the action obtained from BSFT, where the
abelian action given in \cite{MooreBT,MooreBT2}
 was generalised to the non-abelian case. 

The results presented in this paper can be considered 
as an additional
support for the approach given in \cite{Terashima}. More precisely, in this paper we
will study the problem of the tachyon condensation in
system of 
$N$ D-instantons in bosonic and type IIA theory 
\footnote{In the same way we could study unstable
D0-branes in type IIB theory.} in the limit $N\rightarrow \infty$.
We will see that the problem of tachyon condensation is
again extremely simple from the point of view of BSFT. 
In particular, we will calculate the partition sum on the disk, which
is basic ingredient of BSFT,  for
configurations of D-instantons corresponding to higher dimensional
D-branes
(For review of the construction of higher dimensional D-branes
from the lower dimensional ones, see \cite{TaylorM1,TaylorM2,
Schwarz,TaylorM3})  with  the tachyon field turned
on. This approach is similar to the study of the systems of $N$ 
D-instantons in the boundary state formalism
\cite{Ishibashi1,Ishibashi2,OkuyamaBS,Alwis}. 
We hope that our approach can bring new facts about
the tachyon condensation on the system of N D-instantons
and emergence of all D-branes from the lower dimensional
ones. In particular, we hope that this approach could have some relation
to the Matrix theory.

The organisation of the paper is as follows. In section
(\ref{second}) we  give a brief review of BSFT. In section
(\ref{third}) we present the calculation of the partition
function for D-instantons in the bosonic string theory. In
particular, we will show that with the  background
configuration of $N$ D-instantons  well known from the matrix
theory \cite{TaylorM1,TaylorM2} we are able to 
obtain all even dimensional D-branes from the point
of view of BSFT.
Then we extend this approach to the case of odd dimensional
D-branes. However, to find  such a solution we 
should turn on the tachyon mode in the BSFT and
study its condensation in the same way as
in \cite{MooreBT,MooreBT2}.

In section (\ref{fourth}) we will analyse the 
same problem in the superstring BSFT, using
the recent proposal \cite{MooreBT2} relating
BSFT action and the superstring partition sum on
the disk.
We will again see that from the system
of $N$ D-instantons we obtain all D-branes in
type IIA theory.  
In conclusion (\ref{fifth}) we  outline our results
and suggest possible extension of our work.

\section{A brief review of the boundary
string field theory}\label{second}

In this section we review the basic facts about
boundary string field theory 
\cite{WittenBT}. A  general string field configuration
in the boundary string field theory
(or background independent string field theory, in any
case BSFT) is associated with a boundary operator
of ghost number one in the world-sheet theory of
matter and ghost system. 
We will take the world-sheet to be a disk with unit
radius and the metric
\begin{equation}
ds^2=d\rho^2+\rho^2d\tau^2, 0<\rho<1, \
\tau\in <0, 2\pi > \ .
\end{equation}
If  $\left\{\mathcal{O}_I\right\}$ denotes a set
of boundary vertex operators of ghost number 1, 
we can expend a general operator $
\mathcal{O}$  of ghost number 1 as
\begin{equation}
\mathcal{O}=\sum_I \lambda^I\mathcal{O}_I \ .
\end{equation}
We shall restrict to the operators of the form
$\mathcal{O}=c\mathcal{V}=\sum_Ic\lambda^I
\mathcal{V}_I$ where $c$ is the ghost field 
and $\mathcal{V}=\sum_I\lambda^I\mathcal{V}_I$
is a boundary operator in matter theory. In this
case, a string field configuration
associated with the operator $c\mathcal{V}$ is
described by the world-sheet action
\begin{equation}\label{Sworld}
S=S_{Bulk}+\int_0^{2\pi}\frac{d\tau}{2\pi}
\mathcal{V}(\tau) \ ,
\end{equation}
where $S_{Bulk}$ denotes the bulk world-sheet
action corresponding to the closed string background. 
When we  restrict to the trivial  background,
then $S_{Bulk}$ describes of conformal field
theory (CFT) of  free 26 scalar fields $X^{\mu}$
\footnote{In this paper we consider the space-time
with Euclidean signature and the metric
$g_{\mu\nu}=\delta_{\mu\nu}$.}
 and  the $(b,c)$ ghost system. 
For non-abelian case, the boundary interaction 
$\mathcal{V}$ is promoted to $N\times N$ matrix and
the path integral measure on the disk is weighted with
\begin{equation}
e^{-S_{Bulk}}\tr P\exp\left(-\int_0^{2\pi}
\frac{d\tau}{2\pi}\mathcal{V}\right) \ .
\end{equation}
For such configurations, the bosonic string field theory action
$S(\lambda^I)$ is obtained as a solution of
equation
\begin{equation}\label{def}
\frac{\partial S}{\partial \lambda^I}
=\frac{K}{2}\int \frac{d\tau}{2\pi}
\int \frac{d\tau'}{2\pi}
\left<\mathcal{O}_I(\tau) 
\{Q_B,\mathcal{O}(\tau')\}\right>_{
\mathcal{V}} \ ,
\end{equation}
where $<\dots>_{\mathcal{V}}$ denotes
 correlation function in the world-sheet
field theory described by the action
(\ref{Sworld}), $Q_B$ is BRST charge and
$K$ is normalisation constant which has
been fixed in \cite{SenBT} as
\begin{equation}
K=T_p \ , T_p=\frac{2\pi}{(4\pi^2\alpha')
^{(p+1)/2}} \ .
\end{equation}
It is clear that (\ref{def}) determines  action
up to additive constant. 

It was also shown in \cite{WittenBT,WittenBT1,
ShatasviliBT,ShatasviliBT1} that the action
(\ref{def}) is related to the partition sum on
the disk $Z(\lambda)$ via
\begin{equation}
S=\left(\sum_I \beta^I\frac{\partial}{\partial 
\lambda^I}+1\right)Z(\lambda) \ .
\end{equation}
This definition also fix the ambiguity in the
definition of the action by the requirement that
at the fix points of the boundary RG (at
which $\beta(\lambda^*)=0$) 
\begin{equation}
S(\lambda^*)=Z(\lambda^*) \ .
\end{equation}
These results corresponds to abelian $U(1)$
case. The generalisation to the non-abelian
case seams to be nontrivial task thanks to
the presence of the path-ordered product in
the boundary interaction terms. Instead of
address this problem in this paper, we will
proceed in slightly different way. The main
point of our calculation will be the partition sum
on the disk that, as we have seen, plays the prominent
role in the abelian case. We will calculate the
partition sum for different configurations
of D-instantons and we will show that 
 it corresponds to the partition sum
for specific D-brane in particular background.
Then we use  well known results from the
abelian case to obtain the correct tension of
the resulting D-brane. The correctness of
our analysis will be especially seen in the
supersymmetric case where it was conjectured
\cite{MooreBT2} that the partition sum
is equal to the space-time action. It would
be certainly nice to obtain non-abelian action
directly from BSFT theory, especially in
bosonic case. We hope to return to
this problem in the future.

In the next section we will calculate the
partition sum in the bosonic theory for
$N$ D-instantons and we will show that 
from this configuration all D-branes in
bosonic string theory naturally emerge.

\section{D-instantons in BSFT}\label{third}
In this section we would like to show that
through the configuration of $N$ D(-1)-branes
(D-instantons)
in the limit $N\rightarrow \infty$  it is
possible to obtain all D-branes of even and odd dimensions
from the BSFT theory. 
We begin with the construction  of 
even dimensional D-branes in bosonic string theory.
 The starting point is the evaluation
of the  partition sum on the disk 
\begin{equation}\label{bosB}
Z=<\tr P\exp\left(i\int_{0}^{2\pi} d\tau
\Phi_I\partial_n X^I(\tau) \right)> \ ,
\end{equation}
where $\tau$ is a coordinate labelling
the  boundary of the disk and
$P$ is the  path-ordering defined as
\begin{equation}
Pe^{\int d\tau M(\tau)}=
\sum_{N=0}^{\infty}
\int d\tau^1\dots d\tau^N \theta (\tau_{12})\dots
\theta(\tau_{N-1,N})M(\tau_1)\dots M(\tau_N) , \
 \theta(\tau_{12})=\theta(\tau_1-\tau_2) \ .
\end{equation}
In (\ref{bosB}) $\Phi_I, \ I=1,\dots,26$ are Hermitean 
matrices describing the
background configuration of $N$ D-instantons and $X^I(\tau)$
is a string field living on the boundary of the world-sheet
and $\partial_n$ is a normal derivative to the boundary of
the world-sheet.
 In the following
we will always presume the limit $N\rightarrow \infty$.

 Let
us consider the background configuration
\begin{equation}\label{backk}
\Phi_a=\Phi_a,  \ [\Phi_a,\Phi_b]=i\theta_{ab} , 
\ a,b=1,\dots, 2p, \ 
\Phi_{\alpha}=0, \ \alpha=2p+1,\dots, 26 \ .
\end{equation}  
We will show that this configuration corresponds to D(2p-1)-brane 
siting in the origin of the transverse space labelled with coordinates
$x^{\alpha}, \ \alpha=2p+1,\dots, 26$ and with the 
the gauge field background turned on its world-volume
 with the constant field strength
$F_{ab}=\theta_{ab}$.

From (\ref{backk}) we see that $\Phi_a$ are equivalent to
the quantum mechanics operators with nontrivial commuting
relations. Then we can use the well known relation between
the trace over Hilbert space and the path integral
\begin{eqnarray}
\tr P\exp\left (-i\int d\tau H(\tau)\right)=
\int \bra{q}\exp \left(-i\int d\tau H(\tau) \right)
\ket{q}= \nonumber \\
=\int [dq][dp] 
\exp\left( i\int d\tau\left[p(\tau) \dot{q}(\tau) -H(p(\tau),q
(\tau)) \right]\right) \ , \nonumber \\
\end{eqnarray}
with $\dot{q}=\partial_{\tau} q$ and where $p$ is a momentum
conjugate to $q$. In our case, the Hamiltonian is
\begin{equation}
H(\tau)=-\Phi_a\partial_nX^a(\tau) \ .
\end{equation}
with the  operators $\Phi_a$. Then
we can rewrite the trace in (\ref{bosB}) as a path integral
\begin{equation}
\int \prod_{a=1}^{2p}
[\phi_a]\exp\left(i\int d\tau
(\frac{1}{2}\phi_a(\tau)\theta^{ab}\dot{\phi}_b(\tau)
+\phi_a(\tau)\partial_nX^a (\tau))\right) \  ,
\end{equation}
with $\theta_{ac}\theta^{cb}=\delta_a^b$.
We can easily perform the integration over $\phi$.
Firstly, the kinetic term is equal to
\begin{eqnarray}
\frac{i}{2}\int d\tau \phi_a(\tau)\theta^{ab}
\dot{\phi}_b(\tau)=
\frac{1}{2}\int d\tau d\tau'\phi_a(\tau')i\delta (
\tau'-\tau)\theta^{ab}\dot{\phi}_b(\tau)=\nonumber \\
=-\frac{1}{2}\int d\tau d\tau' 
\phi_a(\tau')i\partial \delta(\tau'-\tau)
\theta^{ab}\phi_b(\tau) \ , \partial=\partial_{\tau}, \
\partial'=\partial_{\tau'} \ , \nonumber \\
\end{eqnarray}
so that  the path integral is equal to (up to 
numerical factor)
\begin{eqnarray}\label{Jef}
\int [d\phi^a]\exp\left(-\int d\tau d\tau' \left[
\frac{1}{2}\phi_a(\tau')\triangle(\tau',\tau)^{ab}
\phi_b(\tau)-\phi_a(\tau)
i\partial_n X^a(\tau)\right]\right)= \nonumber \\
=\exp\left(-\frac{1}{2}\int d\tau d\tau'
\partial_nX^a(\tau)\triangle(\tau,
\tau')_{ab}^{-1}\partial_n X^b(\tau') \right) \ ,
\triangle (\tau',\tau)^{ab}=i\partial 
\delta (\tau'-\tau)\theta^{ab} \ . \nonumber \\
\end{eqnarray}
To proceed further
 we must calculate $\triangle^{-1}$ that
should obey
\begin{equation}
\int_0^{2\pi} d\tau''\triangle^{-1}(\tau-\tau'')
\triangle (\tau''-\tau')=\delta (\tau-\tau')-\frac{1}{2\pi} ,  \ 
\delta(\tau-\tau')=\frac{1}{2\pi}\sum_n
e^{in(\tau-\tau')} \ . 
\end{equation}
From this definition we easily obtain
\begin{equation}
\triangle(\tau',\tau)^{ab}=\theta^{ab}
\frac{1}{2\pi}
\sum_n ne^{in(\tau'-\tau)}
\Rightarrow
\triangle (\tau,\tau')^{-1}_{ab}=
\theta_{ab}\frac{1}{2\pi}
\sum_n\frac{1}{n}e^{in(\tau-\tau')} \ .
\end{equation}
We expand the string field as 
\begin{equation}
X^a(\tau,\rho)=\sqrt{
\frac{\alpha'}{2}}\sum_{n=-\infty,
n\neq 0}^{\infty}\rho^n X^a_ne^{in\tau}, \
\partial_n X^a(\tau,\rho=1)=
\sqrt{\frac{\alpha'}{2}}\sum_{n=-\infty, n\neq 0}^{\infty}
n X^a_n e^{in\tau} \ .
\end{equation}
Note that there is no zero mode thanks to
the Dirichlet boundary conditions.
Then (\ref{Jef}) is equal to
\begin{eqnarray}
-\frac{1}{2}\int d\tau d\tau'
\partial_nX^a(\tau) \triangle(
\tau,\tau')_{ab}^{-1}\partial_n X^b(\tau')=
\alpha'\pi\sum_{m=1}^{\infty}
m \theta_{ab}X^a_{-m}X^b_{m}\ , \nonumber \\
\end{eqnarray}
that  agrees precisely with the expression
\cite{KrausBST}
\begin{equation}
S=\frac{i}{2}\int_0^{2\pi} d\tau F_{ab}X^a(\tau)\dot{X}^b(\tau)=
\pi\alpha'\sum_{n=1}^{\infty}nF_{ab}
X^a_{-n}X^b_n \ , F_{ab}=\theta_{ab} \ ,
\end{equation}
 arising from the term
\begin{equation}
S=-i\int_0^{2\pi} d\tau A_a(X^a)\partial_{\tau}X^a(\tau) \ .
\end{equation}
in case of the constant field strength.
As a result, we have obtained the partition sum on the disk
for a D(2p-1)-brane in the presence of the
gauge field $F_{ab}$. Since the gauge field corresponds
to the marginal operator, the partition sum is equal
to space-time action that is the Dirac-Born-Infeld
action \cite{TseytlinSM,TseytlinDBI} 
\begin{equation}
S=T_{2p-1}\int d^{2p}x\sqrt{\det (\delta_{ab}
+2\pi\alpha' F_{ab})} \ .
\end{equation}
We must mention that the construction
presented above is an analogue
of the construction of even dimensional D-branes
 from the low-energy 
action for $N$ D-instantons 
\cite{Seiberg,Kluson1} based on the Matrix
theory (For nice review, see \cite{TaylorM1,TaylorM2}).

In order to describe  odd dimensional D-brane we should include
the tachyon into the boundary interaction 
term for $N$ D-instantons and study its tachyon condensation.
Let us include this matrix valued tachyon field. In this case 
the boundary
action has a form
\begin{equation}
S_{bound}=\frac{1}{2\pi}\int d\tau T-
i\int d\tau \Phi_a\partial_nX^a(\tau)=
-i\int d\tau\left(\frac{i}{2\pi}T+
\Phi_a\partial_nX^a(\tau)\right) \ ,
\end{equation}
where $T$ does not depend on 
the world-sheet field $X(\tau)$ as
it should be for D-instanton background. 
In order to obtain D(2p-2)-brane we 
 propose the tachyon field in the form
\begin{equation}
T=u(\Phi_2)^2 \ , 
\end{equation}
where $\Phi$ are given as in (\ref{backk}). For simplicity
we consider  the matrix $\theta_{ab}$ in the
form
\begin{equation}
\theta_{ab}=\left(\begin{array}{ccc}
0 & \theta & 0 \\
-\theta & 0 & 0 \\
0& 0  & \theta_{ij} \\ \end{array}\right) \ , i, j=3,\dots,2p \ .
\end{equation}
In this case the partition sum is equal to
\begin{eqnarray}
Z=<e^{-S_{bound}}>=<\tr P\exp \left(i\int d\tau
\left[\frac{i}{2\pi}u(\Phi_2)^2+\Phi_a\partial_nX^a(\tau)
\right]\right)>=\nonumber \\
=<\tr P\exp(-i\int_{0}^{2\pi}d\tau H(\tau))> \ . \nonumber \\
\end{eqnarray}
As in the case of pure gauge field 
we rewrite this expression using the
path integral formalism. 
The factor $\triangle(\tau-\tau')^{ij}$
is the same as in the case of the boundary
interaction 
without tachyon studied above  and for $x,y=1,2$ we
obtain from the expression
\begin{eqnarray}
i\int d\tau L(\tau)=i\int
d\tau \left( \frac{1}{2}\phi_x(\tau)\theta^{xy}
\dot{\phi}_y(\tau)+\frac{i}{2\pi}u(\Phi_2(\tau))^2+
\partial_nX^x(\tau)\phi_x(\tau)\right)=\nonumber \\
=-\int d\tau d\tau' \left(\frac{1}{2}
\phi_x(\tau')\triangle (\tau',\tau)^{xy}\phi_y(\tau)-
i\phi_x(\tau)\partial_n X^x(\tau)\right) \ ,
x,y=1,2 \ , \nonumber \\
\triangle (\tau',\tau)^{xy}=i\partial \delta(
\tau'-\tau)\theta^{xy}-\frac{1}{\pi}u\delta (\tau'-\tau)
\delta^{2x} \ , \nonumber \\ 
\end{eqnarray}
or more precisely
\begin{eqnarray}
\triangle (\tau'-\tau)^{xy}=\frac{1}{2\pi}
\sum_n \left(n\theta^{xy}-
\frac{u}{\pi}\delta^{2x}\right)e^{in(\tau-\tau')}=
\sum_{n} E^{xy}_n e^{in(\tau-\tau')} \ , \nonumber \\
E^{xy}_n=\left(\begin{array}{cc}
0 & \frac{1}{\theta} n \\
-\frac{1}{\theta} n & -\frac{1}{\pi}u \\ \end{array}\right) \ ,
(E_n)_{xy}^{-1}=\frac{\theta^2}{n^2}\left(
\begin{array}{cc} -\frac{1}{\pi}u & -\frac{n}{\theta} \\
\frac{n}{\theta} & 0 \\ \end{array}\right) \ , \nonumber \\
\triangle^{-1}(\tau-\tau')_{xy}=
\frac{1}{2\pi}\sum_m(E_m)^{-1}_{xy}e^{im(
\tau-\tau')}  \ ,  \nonumber \\
\end{eqnarray}
so that  the partition sum corresponding to $X^x, \ x=1,2$
 is equal to
\begin{equation}
Z=<\exp \left(-\frac{1}{2}\int d\tau d\tau'
\partial_nX^x(\tau)\triangle^{-1}(\tau-\tau')_{xy}
\partial_nX^y(\tau')\right) > \ ,
\end{equation}
where we have
\begin{equation} 
-\frac{1}{2}\int d\tau d\tau'
\partial_nX^x(\tau)\triangle^{-1}(\tau-\tau')_{xy}
\partial_nX^y(\tau')=
\frac{\alpha'\pi}{2}\sum_{
m=-\infty, m\neq 0}^{\infty}\left(m\theta_{xy}X^x_{-m}X^y_m
-\frac{\theta^2}{\pi}X^1_{-m}uX^1_m\right) \ . 
\end{equation}
We see that the second  term above 
corresponds to
\begin{equation}
-\frac{\alpha'}{2}u\sum_{n=-\infty,
n\neq 0}^{\infty} X_{-n}^1X_n^1=
-\frac{1}{2\pi}\int_0^{2\pi} d\tau uX^1(\tau)
X^1(\tau) \ ,
\end{equation}
where we have made a substitution
\begin{equation}
u\theta^2\rightarrow u \ ,
\end{equation}
and where we have used the fact that in the original 
instanton model the zero modes $X^a_0$ are missing. 
When we combine the previous expression with
the  gauge field term $A_i \partial_{\tau}X^i(\tau)$
whose emergence is the same as in the pure gauge field case,
we obtain the partition sum on the disk in the 
well known form 
\begin{equation}\label{gandt}
Z=<e^{-S_{bound}}>=
<\exp\left(-\int_0^{2\pi}
 d\tau \left[\frac{1}{2\pi}u(X^1(\tau))^2-iA_a(X(\tau))
\partial_{\tau}X^a(\tau)\right]\right)> \ .
\end{equation}
Now it is easy to find the value of the action
for the resulting object, following the recent results
\cite{MooreBT,Okuyama,Aleksejev}. For our purpose we
will closely follow \cite{Aleksejev}. We must also stress
that thanks to the  form of the boundary
interaction (\ref{gandt}) which corresponds to
 D(2p-1)-brane with the background gauge field, we should
include the zero mode of the string field $X^a(\tau)$ into
our analysis since we have effectively replaced the
Dirichlet boundary conditions with the boundary conditions
appropriate for the study of the tachyon condensation on
the D(2p-1)-brane.

It is clear that the gauge field $F_{ij}$ decouples from
the tachyon term and consequently contributes to the 
partition function with the factor
\begin{equation}
\int d^{2p-2}x\sqrt{\det (\delta_{ij}+2\pi\alpha' F_{ij})} \ .
\end{equation}
The contribution from the $x,y=1,2$ sector is almost the same
as in \cite{Aleksejev}
\begin{equation}
\int dx^2\sqrt{v}e^{\gamma v} e^{-a}\Gamma(v) \ ,
v=\frac{1}{1+(2\pi\alpha' \theta)^2}2\alpha' u \ ,
\end{equation}
where the integral over $x^2$ is a consequence of
the existence of the  zero
mode in $X^2(\tau)$. In the previous expression we have also included
the constant part of the tachyon $a$ which would correspond
in the original matrix model to the term $a1_{N\times N}$.
The whole partition function is equal to
\begin{equation}\label{Zu}
Z(v,a,F)=Ke^{-a}\sqrt{v}e^{\gamma v}\Gamma(v)
\int d^{2p-1}x\sqrt{
\det (\delta_{ij}+2\pi\alpha' F_{ij})} \ ,
\end{equation}
where $K$ is a  numerical factor which will be determined
in a moment. As was shown in \cite{WittenBT2,MooreBT,Okuyama,
Aleksejev} the space-time action is then equal to
\begin{equation}
S(a,v,F)=(v-a\frac{\partial }{\partial a}-
v\frac{\partial}{\partial v}+1)Z(a,v,F) \ ,
Z(v)=\sqrt{v}e^{\gamma v}\Gamma(v) \ .
\end{equation}
Using this equation we  determine the normalisation
factor $K$ as in \cite{MooreBT}. We can expect that
for $u\rightarrow 0$ this action should reduce to
the DBI action  with the gauge field $F$ and with the
tachyon potential  
\begin{equation}
S(T,F)=T_{2p-1}\int d^{2p}xe^{-T}
\sqrt{\det(\delta_{ab}+2\pi\alpha'F_{ab})}
+O(\partial T) \ ,
T_{2p-1}=\frac{2\pi}{(4\pi^2\alpha')^p}  \ .
\end{equation}
In order to determine the factor $K$ we evaluate this action
on  the tachyon profile $T=a+u(x^1)^2$ and compare
this result with the partition function (\ref{Zu}) in
the limit $u\rightarrow 0$. The previous
expression gives
\begin{eqnarray}
S(T,F)=T_{2p-1}\int d^{2p-1}x\sqrt{\det (\delta_{ab}
+2\pi\alpha' F_{ab})}e^{-a}
\int e^{-u(x^1)^2}dx^1=\nonumber \\
=T_{2p-1}\int d^{2p-1}x
e^{-a}\sqrt{\det(\delta_{ab}+2\pi\alpha'F_{ab}
)}\sqrt{\frac{\pi}{u}}
 \ , \nonumber \\
\end{eqnarray}
and when we compare this result with (\ref{Zu}) evaluated
for  small u we obtain
\begin{eqnarray}
Z(a,v,F)= K\int d^{2p-1}xe^{-a}
\sqrt{\det(\delta_{ij}+2\pi\alpha'F_{ij})}
\frac{1}{\sqrt{v}}=\nonumber \\
=K\int d^{2p-1}xe^{-a}\sqrt{\det (\delta_{ij}+
2\pi\alpha'F_{ij})}\sqrt{1+(2\pi\alpha'\theta)^2}
\frac{1}{\sqrt{2\alpha'u}} \Rightarrow K=T_{2p-1}
\sqrt{2\pi\alpha'} \ . \nonumber \\
\end{eqnarray}
 Now we are ready to calculate
the tension of the resulting D(2p-2)-brane, following
\cite{MooreBT}.
Variation of  this action with respect to $a$  gives
\begin{equation}
\frac{\delta S}{\delta a}=
[v+a-v\frac{\partial}{\partial v}]Z(v)=0\Rightarrow
a^{\star}=-v+v\frac{d \ln Z(v)}{dv} \ ,
\end{equation}
and the action evaluated with this value of $a^*$
is
\begin{equation}
S(a^{\star},v,F)
=\exp(\Theta(v))K\int d^{2p-1}x\sqrt{\det
(\delta_{ij}+2\pi\alpha' F_{ij})} , \ 
\Theta(x)=x-x\frac{d \ln Z(x)}{dx}
+\ln Z(x)  \  . 
\end{equation}
 As was shown in \cite{WittenBT1,MooreBT} 
the minimum  of $\Theta(x)$
corresponds to $x$ going to 
infinity. In this limit we have \cite{MooreBT} 
\begin{equation}
\Theta(x)=\log \sqrt{2\pi}+\frac{1}{6x}+
O(\frac{1}{x^2}) \ , x\rightarrow \infty \ ,
\end{equation}
which gives the value of the action 
\begin{equation}
S=T_{2p-1}\sqrt{4\pi^2\alpha'}\int
d^{2p-1}x
\sqrt{\det (\delta_{ij}+2\pi\alpha'F_{ij})} \ .
\end{equation}
With using $T_{2p-1}\sqrt{4\pi^2\alpha'}=T_{2p-2}$
we see that the previous expression is an exact value of
the action for D(2p-2)-brane with the background
gauge field $F_{ij}$. Note that 
this action does not depend on 
the gauge field $F_{xy}$.
 
We have seen in this section  that
from the configurations of $N$
 D-instantons in the limit $N\rightarrow
\infty$ we obtain 
all D-branes in the bosonic string  theory.
We have analysed  this system in
the framework of BSFT
which again leads to the correct result as in
\cite{MooreBT}. 
In the next section we extend this analysis to
the case of non-BPS and BPS D-branes in
type IIA theory. Extension to the type IIB theory
is straightforward, we simply start with the
configuration of unstable D0-branes.

\section{D-instantons in the superstring boundary
string field theory}\label{fourth}
In this section we would like to analyse the same problem
in the context of the superstring boundary
string field theory.  In this case the calculation is 
simpler since as was conjectured in \cite{MooreBT2}
\footnote{Similar  approach to this problem has also
been  discussed in \cite{TseytlinSM} where 
many  references on the relevant papers can be found.}  the
space-time action $S$ is equal to the disk partition
sum
\begin{equation}
S(\lambda_I)=Z(\lambda_I) \ ,
Z(\lambda_I)=<e^{-S_{bound}(\lambda_I)}> \ .
\end{equation}
 The bulk action is
\begin{equation}
S_{Bulk}=\frac{1}{4\pi}\int d^2z
\left(\frac{2}{\alpha'}\partial X^{\mu}
\partial X_{\mu}+\psi^{\mu}\overline{\partial}
\psi_{\mu}+\tilde{\psi}^{\mu}
\partial\tilde{\psi}_{\mu}\right) \ .
\end{equation}
The mode expansion at the boundary of the disk
is \cite{KrausBST} (in NS sector)
\begin{equation}
\psi^{\mu}(\tau)=\sum_{r=1/2}^{\infty}
\left(\psi^{\mu}_re^{ir\tau}+\psi_{-r}^{\mu}e^{-ir\tau}\right) \ .
\end{equation}
and the bulk action
\begin{equation}
S_{Bulk}=\frac{1}{2}\sum_{n=1}^{\infty}
nX_{-n}^{\mu}X_{n}^{\nu}\delta_{\mu\nu}+
i\sum_{r=1/2}^{\infty}r\psi^{\mu}_{-r}\psi^{\nu}_r 
\delta_{\mu\nu},  \ \mu,\nu=1,\dots, 10 \ .
\end{equation}
The  next step is to introduce the boundary interactions which
should be invariant under supersymmetry transformations. This
can be done by working in boundary superspace 
\cite{MooreBT2,KrausBST,TakayBT,Andrejev,TseytlinSM}
 with coordinates
$\hat{\tau}=(\tau,\theta)$ and
\begin{equation}
{\bf X}^{\mu}=X^{\mu}+\sqrt{\alpha'}\theta
\psi^{\mu}, \ 
D=\partial_{\theta}+\theta\partial_{\tau} \ .
\end{equation}
The simplest example of supersymmetric boundary action
corresponds to gauge field. For abelian gauge field the
supersymmetric boundary action is
\begin{eqnarray}
-i\int d\tau d\theta A_{\mu}({\bf X})D{\bf X}^{\mu}=
-i\int d\tau d\theta (A_{\mu}(X)+\partial_{\nu}A_{\mu}(X)\sqrt{\alpha'}
\theta\psi^{\nu})(\sqrt{\alpha'}\psi^{\mu}+\theta \partial_{\tau}
X^{\mu})=\nonumber \\
=-i\int d\tau \left( A_{\mu}(X(\tau))\partial_{\tau}X^{\mu}
+\frac{1}{2}\alpha'F_{\mu\nu}\psi^{\mu}\psi^{\nu} \right)
, \ \nonumber \\
\end{eqnarray}
where we have used 
\begin{equation}
\theta^2=0, \ \int d\theta=0, \ \int d\theta \theta=1 \ .
\end{equation}
In case of Dirichlet boundary conditions for $x^a, \ a,b=p+1,\dots,10$
 and
Neumann boundary conditions for $x^{\mu}, \mu,\nu=1,\dots, p$
 we will write the boundary action as
\begin{eqnarray}
-i\int d\tau d\theta \left (A_{\mu}({\bf X}^{\mu})D
{\bf X}^{\mu}+\Phi_a({\bf X^{\mu}})\tilde{D}{\bf X}^a\right)=
\nonumber \\
-i\int d\tau \left( A_{\mu}(X^{\mu}(\tau))\partial_{\tau}X^{\mu}(\tau)
+\frac{1}{2}\alpha'F_{\mu\nu}\psi^{\mu}(\tau)\psi^{\nu}(\tau) \right)
-\nonumber \\ -i
\int d\tau\left(\Phi_a(X^{\mu}(\tau))\partial_nX^a(\tau)+\alpha'
\partial_{\mu}\Phi_a(X^{\mu}(\tau))\psi^{\mu}(\tau)\psi^a(\tau) \right) \ ,
\nonumber \\
\end{eqnarray}
where we have used the fact that the background fields are
functions of the string  fields $X^a(\tau)$
 obeying Neumann boundary conditions
and we have defined
\begin{equation}
\tilde{D}=\partial_{\theta}+\theta\partial_n \ . 
\end{equation}
Non-abelian extension is given (In case of all Neumann boundary 
conditions, generalisation to the case of lower dimensional D-branes
is straightforward as we will see in case of D-instantons.)
\begin{equation}
e^{-S_A}=\tr \hat{P}e^{i\int d\tau d\theta A_{\mu}({\bf X})
D{\bf X}^{\mu}} \ ,
\end{equation}
where the symbol $\hat{P}$ is defined as
\begin{equation}
\hat{P}e^{\int d\hat{\tau} M(\hat{\tau})}=
\sum_{N=0}^{\infty}
\int d\hat{\tau}_1\dots d\hat{\tau}_N
\Theta(\hat{\tau}_{12})\Theta
(\hat{\tau}_{23})\dots \Theta(\hat{\tau}_{N-1,N})
M(\hat{\tau}_1)M(\hat{\tau_2})\dots
M(\hat{\tau}_N) \  ,
\end{equation}
where $d\hat{\tau}=
d\tau d\theta, \ \hat{\tau}_{12}=\tau_1-\tau_2-\theta_1\theta_2$
and $\Theta$ is a step function whose expansion is
equal to $\Theta(\hat{\tau}_1-\hat{\tau}_2)=
\theta(\tau_1-\tau_2)-\delta(\tau_1-\tau_2)
\theta_1\theta_2$. As was shown in \cite{Andrejev},
these contact terms are essential for world-sheet
supersymmetry and they are also 
crucial for gauge invariance as they contribute with
the $[A_{\mu},A_{\nu}]$ in the field strength. More precisely, 
the integration
over $\theta$ gives the result
\begin{equation}
e^{-S_A}=\tr Pe^{i\int d\tau [A_{\mu}(X)\partial_{\tau}A^{\mu}+
\frac{\alpha'}{2}F_{\mu\nu}\psi^{\mu}\psi^{\nu}]} \ ,
F_{\mu\nu}=\partial_{\mu}A_{\nu}-
\partial_{\nu}A_{\mu}-i[A_{\mu},A_{\nu}] \ .
\end{equation}
It is easy to generalise this boundary interaction to the case of
lower dimensional  Dp-branes.
We will mainly consider D-instantons and for them the boundary interaction
is
\begin{equation}\label{SA}
e^{-S_A}=\tr \hat{P}\exp \left(i\int d\hat{\tau}\Phi_I\tilde{
D}{\bf X}^I \right)=\tr P\exp\left(
i\int d\tau \left[\Phi_I\partial_nX^I(\tau)-i\frac{\alpha'}{2}
[\Phi_I,\Phi_J]\psi^I(\tau)\psi^J(\tau)\right]\right) \  ,
\end{equation}
with $I,J=1,\dots, 10$.
As in bosonic case, the description of
 even dimensional D-branes is very straightforward
and corresponds to the  natural emergence of
these D-branes in the matrix models \cite{Seiberg,Kluson1,Terashima,
TaylorM1,TaylorM2}. 
Let us consider the background configuration
in the form
\begin{equation}
[\Phi_a,\Phi_b]=i\theta_{ab}, \ a=1,\dots, 2p, \ 
\Phi_{\alpha}=0, \ \alpha=2p+1,\dots, 10 \ .
\end{equation}
Then the fermionic term  in 
(\ref{SA}) is equal to
\begin{equation}
-i\frac{\alpha'}{2}[\Phi_a,\Phi_b]\psi^a(\tau)
\psi^b(\tau)=\frac{\alpha'}{2}\theta_{ab}\psi^a(\tau)
\psi^b(\tau)1_{N\times N} \ .
\end{equation}
We see that this term is proportional to the unit matrix so that
the following expression can be taken out the trace
\begin{equation}
\exp\left( i\int d\tau \frac{\alpha'}{2}\theta_{ab}\psi^a 
(\tau)\psi^b(\tau)\right) \ .
\end{equation}
Now it is easy to see, following the calculations
presented in section (\ref{third}), that this D-instantons
configurations describe D(2p-1)-brane with 
 the background gauge
field $F_{ab}=\theta_{ab}$  with the boundary action
\begin{equation}
e^{-S_A}=\exp \left(i\int d\tau\left[
A_a(X)\partial_{\tau}X^a(\tau)+\frac{\alpha'}{2}
F_{ab}\psi^a(\tau)\psi^b(\tau)\right]\right), 
F_{ab}=\partial_aA_b-\partial_bA_a=\theta_{ab} \ .
\end{equation}
Consequently the action evaluated on this particular
D-instanton background is  equal to
\begin{equation}
S(F)=Z(F)=<e^{-S_A}>=\sqrt{2}T_{2p-1}
\int d^{2p}x\sqrt{\det (\delta_{ab}+2\pi\alpha'F_{ab})} \ .
\end{equation}

In order to describe  odd dimensional
D-brane we should include the tachyon
in the boundary action and calculate its condensation.
The inclusion of tachyon into the boundary
interaction in supersymmetric case has been done in
\cite{MooreBT2,KrausBST,TseytlinSM}. We introduce superfield
$\Gamma=\xi+\theta F$ living on the boundary
of the disk. Now the tachyon boundary action
is (For all  ${\bf X}(\tau)$ obeying Neumann boundary
conditions)
\begin{equation}
e^{-S_{bound}}=\tr \hat{P}
\exp\left (\int\frac{d\hat{\tau}}{2\pi}
(\Gamma D\Gamma+T({\bf X}))\Gamma \right) \ .
\end{equation}
In abelian case we can easily perform the
integration over $\theta$. When we also
integrate out the auxiliary field
 $\Gamma $ we obtain  \cite{MooreBT2}
\begin{equation}
e^{-S_{bound}}=\exp\left(
-\frac{1}{4}\int \frac{d\tau}{2\pi}\left[
T(X)^2+(\psi^{\mu}\partial_{\mu}T)
\frac{1}{\partial_{\tau}}(\psi^{\nu}
\partial_{\nu}T)\right]\right) \ .
\end{equation}
The generalisation of this action to
 the case of $N$ D-branes with
Neumann boundary conditions in $x^a, a=1,\dots, k$ and
Dirichlet boundary conditions in $x^i, i =k+1,\dots,10$ is
\begin{eqnarray}\label{Z}
S(\Phi,T)=Z(\Phi,T)=
<e^{-S_{A}-S_{bound}}>=\nonumber \\
=<\tr \hat{P}\exp\left(\int d\hat{\tau}\left[
\frac{1}{2\pi}\Gamma D\Gamma+\frac{1}{2\pi}
T({\bf X}^a)\Gamma+iA_{a}({\bf X}^{a})D{\bf X}^a+
i\Phi_i({\bf X}^a)\tilde{D}X^i\right]\right)> \ .
\nonumber \\
\end{eqnarray}
In the following we will consider $N$ D(-1)-branes with
 the boundary interaction 
\begin{equation}
e^{-S_{bound}-S_A}=
\tr\hat{P}\exp\left(\int d\hat{\tau}
\left[\frac{1}{2\pi}\Gamma D\Gamma+
\frac{1}{2\pi}T\Gamma+i\Phi_I\tilde{D}{\bf X}^I
\right]\right), \ I=1,\dots,10 \ .
\end{equation}
The basic idea is the same as in
previous section (\ref{third}). 
We start with performing the integration over $\theta$.
 In order do that, we express
the previous expression using the elegant formalism
\cite{TseytlinSM} in which the boundary action has a form
\begin{equation}
e^{-S_{A}-S_{Bound}}=
\int [d\hat{\eta}][d\hat{\overline{\eta}}]
\exp\left(\int d\hat{\tau}\left[
\hat{\overline{\eta}}_bD\hat{\eta}^b+
\frac{1}{2\pi}\Gamma
D\Gamma+\hat{\overline{\eta}}_a[\frac{1}{2\pi}
T^a_b\Gamma+i\Phi^a_{bI}\tilde{D}{\bf X}^I]
\hat{\eta}^b\right]\right) \ ,
\end{equation}
where the fields $\hat{\eta}^a=\eta^a+\theta \chi^a, \
\hat{\overline{\eta}}_a=\overline{\eta}_a+
\theta \chi_a$ transform in antifundamental and fundamental
representation of $U(N)$. The integration over $\theta $ gives
\begin{eqnarray}
S_{bound}+S_A=I=
\int d\tau\left(
\overline{\eta}_a\dot{\eta}^a-\overline{\chi}_a\chi^a+
\frac{1}{2\pi}\xi\dot{\xi}-
\frac{1}{2\pi}F^2+\right. \nonumber \\
+\left. \frac{1}{2\pi}
\left(\overline{\eta}_aT^a_b\xi \chi^b+
\overline{\eta}_aT^a_bF\eta^b-
\overline{\chi}_aT^a_b\xi \eta^b\right)-\right.
\nonumber \\
\left.-i\overline{\eta}_a\Phi^a_{bI}\sqrt{\alpha'}
\psi^I\chi^b-
i\overline{\eta}_a\Phi^a_{bI}\partial_nX^I
\eta^b-i\overline{\chi}_a\Phi^a_{bI}\sqrt{\alpha'}
\psi^I\eta^b
\right) \ , \nonumber \\
\end{eqnarray}
In the previous expression  $a,b=1,\dots, N$ are matrix
indices, $I=1,\dots, 10$ are space-time indices. 
As a next thing we integrate out
 the auxiliary field $\chi^a, \overline{\chi}_a$
which gives
\begin{eqnarray}\label{I1}
I=\int d\tau\left(
\overline{\eta}_a\dot{\eta}^a+
\frac{1}{2\pi}\xi\dot{\xi}-
\frac{1}{2\pi}F^2+\right. \nonumber \\
+\left.\overline{\eta}_a\left[
\frac{i\sqrt{\alpha'}}{2\pi}\xi\psi^I[
\Phi_I,T]^a_b
+\frac{1}{2\pi}T^a_bF
-i\Phi^a_{bI}\partial_nX^I
-\frac{\alpha'}{2}[\Phi_I,\Phi_J]^a_b
\psi^I\psi^J\right]\eta^b\right) \ . \nonumber \\
\end{eqnarray}
We rewrite   the path integral over
$\overline{\eta},\eta$ in (\ref{I1}) as a trace with 
path ordering $P$ so that  we obtain
\begin{eqnarray}
<e^{-S_{bound}-S_A}>=
<\tr P\exp \left(\int d\tau\left[
-\frac{1}{2\pi}\xi\dot{\xi}+
\frac{1}{2\pi}F^2+\right.\right. \nonumber \\
\left.\left.
-\frac{i\sqrt{\alpha'}}{2\pi}\xi\psi^I[
\Phi_I,T]
-\frac{1}{2\pi}TF
+i\Phi_{I}\partial_nX^I
+\frac{\alpha'}{2}[\Phi_I,\Phi_J]
\psi^I\psi^J\right]\right)> \ .\nonumber \\
\end{eqnarray}
Let us consider the ansatz
\begin{equation}
T=u \Phi_2, \ 
[\Phi_x,\Phi_y]=i\epsilon_{xy}\theta, \ , x,y=1,2, \
 [\Phi_i,\Phi_j]=i\theta_{ij}, \
i, j= 3,\dots, 2p, 
\end{equation}
so that $\theta$ in $[\Phi_a,\Phi_b]=i\theta_{ab}$
has a  form
\begin{equation}
\theta=\left(\begin{array}{ccc}
0 & \theta & 0 \\
-\theta & 0 & 0 \\
0 & 0 & \theta_{ij} \\ \end{array}\right)
\ , i, j=3,\dots, 2p \ .
\end{equation}
Then the previous expression is equal to 
\begin{eqnarray}
<e^{-S_{bound}-S_A}>=
<\int [d\phi_a]\exp
\left(\int d\tau\left[ \frac{1}{2}
i\phi_a\theta^{ab}\dot{\phi}_b
-\frac{1}{2\pi}\xi\dot{\xi}+
\frac{1}{2\pi}F^2+\right.\right. \nonumber \\
\left.\left.
+\frac{\sqrt{\alpha'}}{2\pi}\xi\psi^1u\theta
-\frac{1}{2\pi}u\phi_2F
+i\phi_a\partial_nX^a
+\frac{i\alpha'}{2}\theta_{ab}
\psi^a\psi^b\right]\right)> \ . \nonumber \\
\end{eqnarray}
The integration over auxiliary field $F$ gives
\begin{equation}
F(\tau)=\frac{1}{2}u\phi_2(\tau)
\end{equation}
and we have
\begin{eqnarray}
<e^{-S_{bound}-S_A}>=
<\int [d\phi_a]\exp
\left(\int d\tau\left[ i\frac{1}{2}
\phi_a\theta^{ab}\dot{\phi}_b
-\frac{1}{2\pi}\xi\dot{\xi}
+\right.\right. \nonumber \\
\left.\left.
+\frac{\sqrt{\alpha'}}{2\pi}\xi\psi^1u\theta
-\frac{1}{8\pi}u^2(\phi_2)^2
+i\phi_a\partial_nX^a
+\frac{i\alpha'}{2}\theta_{ab}
\psi^a\psi^b\right]\right)> \ . \nonumber \\
\end{eqnarray}
 We can take out the following expression
from the path integral over $\phi$
\begin{equation}
\exp\left(\int d\tau\left[ -\frac{1}{2\pi}\xi\dot{\xi}
+\frac{\sqrt{\alpha'}}{2\pi}
\xi \psi^1u\theta +\frac{i\alpha'}{2}\theta_{ab}
\psi^a\psi^b\right]\right) \ .
\end{equation}
Next calculation is the same as in  previous
section (\ref{third}).
The  $i,j=3,\dots, 2p$ components  
give the
same result as in the case of even dimensional
D-brane without exciting tachyon and
the integration over $x,y=1,2$ gives
\begin{eqnarray}\label{up}
\exp\left(-\frac{1}{2}\int d\tau d\tau'
\partial_nX^x(\tau)
\triangle(\tau-\tau')_{xy}^{-1}\partial_nX^y
(\tau')\right) \ , \nonumber \\
\triangle(\tau-\tau')_{xy}^{-1}=
\sum_n (E_n)_{xy}^{-1}e^{in(\tau-\tau')}, \ 
(E^{-1}_n)_{xy}=\frac{\theta^2}{n^2}\left(
\begin{array}{cc} -\frac{1}{4\pi}u^2 & -\frac{n}{\theta} \\
\frac{n}{\theta} & 0 \\ \end{array}\right) \ , \nonumber \\
\end{eqnarray}
which together with the $F_{ij}$ and  $F_{xy}$ terms
gives the result
\begin{equation}
Z=<\exp\left(\int d\tau\left[ -\frac{1}{2\pi}\xi\dot{\xi}
+\frac{\sqrt{\alpha'}}{2\pi}
\xi \psi^1u
-\frac{1}{
8\pi}u^2 (X^1)^2+iA_a(X^a)\partial_{\tau}X^a
 +\frac{i\alpha'}{2}\theta_{ab}
\psi^a\psi^b\right]\right)> \ ,
\end{equation}
where we have made replacement $u\theta\rightarrow u$.
We can also integrate out the auxiliary field $\xi$ which leads
to
\begin{equation}
\dot{\xi}(\tau)=\frac{u\sqrt{\alpha'}}{2}
\psi^1(\tau)\Rightarrow
\xi(\tau)=\frac{u\sqrt{\alpha'}}{2}
\frac{1}{\partial_{\tau}}\psi^1(\tau) 
\end{equation}
and we obtain the same expression
 as in
\cite{MooreBT2,Aleksejev}
\begin{eqnarray}
<e^{-S_A-S_{bound}}>=
<\exp \left(
\int d\tau\left[-\frac{1}{
8\pi}(u^2 (X^1)^2+
\alpha'u^2\psi^1\frac{1}{\partial_{\tau}}
\psi^1)\right.\right. + \nonumber \\
\left.\left.+iA_a(X^a)\partial_{\tau}X^a+
\frac{i\alpha'}{2}F_{ab}\psi^a\psi^b\right]
\right)> \ ,F_{ab}=\theta_{ab} \ . \nonumber \\
\end{eqnarray}
Then it is easy to see that the tachyon condensation
really leads to the emergence of odd dimensional
D-brane exactly in the same way as in \cite{MooreBT2}.
The partition function is equal to \cite{MooreBT2,
Aleksejev,Yasnov}
\begin{equation}
Z=K\int d^{2p-1}x
Z(a,v)_{fermi}
\sqrt{\det (\delta_{ij}
+2\pi\alpha'F_{ij})} \ ,
v=\frac{\alpha' u^2/2}{1+(2\pi\alpha' \theta)^2}  \ .
\end{equation}
with \cite{MooreBT2} 
\begin{equation}
Z(a,v)_{fermi}=4^v \frac{Z_1(v)^2}{Z_1(2v)} \ ,
Z_1(v)=\sqrt{v}e^{\gamma v}\Gamma(v) \ ,
\end{equation}
 and where $K$ is a numerical factor that will be
determined as in section (\ref{third}).
When we calculate the partition sum for constant
tachyon $T=a1_{N\times N}$ we obtain the exact
tachyon potential $e^{-\frac{a^2}{4}}$ multiplied 
with the DBI term $\sqrt{\det(\delta_{ab}
+2\pi\alpha'F_{ab})}$ arising from the partition sum
calculated in the pure gauge field background.
 Then we can expect that for a slowly
varying tachyon field the action corresponds to the
non-BPS D(2p-1)-brane action is
\begin{equation}\label{Spom}
S=\sqrt{2}T_{2p-1}\int d^{2p}x e^{-T^2/4}\sqrt{
(1+(2\pi\alpha' \theta)^2)\det (\delta_{ij}+
2\pi\alpha' F_{ij})}+O(\partial T) \ ,
\end{equation}
This action  evaluated on the tachyon profile $T=ux^1$ should
be  equal to the partition sum in the limit $u\rightarrow 0$.
 From this requirement we
can determine the overall  normalisation constant in the
partition sum as well as the normalisation of
the tachyon kinetic term \cite{MooreBT2}.
However for our purposes it is sufficient to obtain the
normalisation term $K$ only. Then 
the  action (\ref{Spom}) evaluated on the
tachyon profile  $T(x)=ux^1$ is equal to 
\begin{eqnarray}
\sqrt{2}T_{2p-1}\int d^{2p-1}x
\sqrt{(1+(2\pi\alpha' \theta)^2)\det (\delta_{ij}+
2\pi\alpha' F_{ij})}\int dx^1 e^{-u^2(x^1)^2/4}=\nonumber \\
=2\frac{\sqrt{\pi}}{u}\sqrt{2}T_{2p-1}
\int d^{2p-1}x\sqrt{(1+(2\pi\alpha' \theta)^2)\det (\delta_{ij}+
2\pi\alpha' F_{ij})} \ . \nonumber \\
\end{eqnarray}
On the other hand, in the limit $u\rightarrow 0$ we have
\begin{equation}
Z(v)_{fermi}=\
 \sqrt{\frac{2}{v}}+O(v) , \ v\sim 0 \ ,
\end{equation}
so that the partition function is equal to
\begin{equation}
Z=K\frac{\sqrt{2}\sqrt{2}}{\sqrt{\alpha'u^2}}\int d^{2p-1}x
\sqrt{(1+(2\pi\alpha' \theta)^2)\det (\delta_{ij}+
2\pi\alpha' F_{ij})}   
\end{equation}
and consequently
\begin{equation}
K=\sqrt{2}T_{2p-1}\sqrt{\pi\alpha'} \ .
\end{equation}
Using this result it is easy to see that the action
arising from the tachyon condensation is
equal to (In this case the tachyon condensation
 corresponds to $u\rightarrow \infty$
\cite{MooreBT2})
\begin{eqnarray}\label{Sfinal}
S=Z(\infty)=
\sqrt{2\pi\alpha'}T_{2p-1}\sqrt{2\pi}\int
d^{2p-1}x\sqrt{\det (\delta_{ij}+
2\pi\alpha' F_{ij})}= \nonumber \\
=T_{2p-2}\int d^{2p-1}x
\sqrt{\det(\delta_{ij}+2\pi\alpha'F_{ij})}\ ,
\nonumber \\
\end{eqnarray}
where we have used
\begin{equation}
Z(v)\sim \sqrt{2\pi}+O(v^{-1}) \ , u\rightarrow \infty \ .
\end{equation} 
The result (\ref{Sfinal}) is a correct value of the
action for D(2p-2)-brane with the 
background gauge field strength $F_{ij}$.

In this section we have studied the emergence of 
all D-branes in type IIA theory from the configurations 
of infinite many non-BPS D(-1)-instantons. Exactly
in the same way we could proceed with the infinite
many non-BPS D0-branes in type IIB theory.
\section{Conclusion}\label{fifth}
In this paper we have tried to show that 
all D-branes in bosonic, type IIA and type IIB theory
can emerge from configurations of infinite many
D-instantons, in case of type IIB theory from 
infinite many D0-branes. We have studied this
system from the point of view of BSFT 
and we have shown that we can very easily
obtain correct values of the tensions of all 
D-branes.

However, many open questions remain. Firstly, it 
would be nice to obtain non-abelian action for
D-instantons, or more generally for all non-abelian
D-branes from the BSFT theory. It seams that this
can be easily done in the case of supersymmetric
string theory at least in some special cases
\cite{TseytlinBI,TseytlinDBI}. We hope to return to these
questions in the future. It would be also interesting
to study the D-brane anti-D-brane system in
the framework suggested in this paper, following
the general construction \cite{KrausBST}. We hope
to return to this problem in the forthcoming publication. 
\\
\\

{\bf Acknowledgements}

We would like to thank Rikard von Unge for very helpful 
discussions. This work was supported by the
Czech Ministry of Education under Contract No.
143100006.


\begin{thebibliography}{61}
\bibitem{WittenBT} E. Witten,
\emph{"On background independent open string
field theory,"} \prd{36}{5467}{1992}, \hepth{9208027}.
\bibitem{WittenBT1} E. Witten,
\emph{"Some computations in background 
independent off-shell string theory,"}
\prd{47}{3405}{1993}, \hepth{9210065}.
\bibitem{WittenBT2} K. Li and E. Witten,
\emph{"Role of short distance behaviour in
off-shell open string field theory,"} 
\prd{48}{853}{1993}, \hepth{9303067}.
\bibitem{ShatasviliBT} S. L. Shatashvili,
\emph{"Comment on the background independent 
open string theory,"} \plb{311}{83}{1993},
\hepth{9303143}.
\bibitem{ShatasviliBT1} S. L. Shatashvili,
\emph{"On the problems with background independence
in string theory,"} \hepth{9311177}.
\bibitem{ShatasviliBT2} A. A. Gerasimov and
S. L. Shatashvili, 
\emph{"On exact tachyon potential in
open string field theory,"}, \jhep{0010}{034}{2000},
 \hepth{0009103}.
\bibitem{MooreBT} D. Kutasov, M. Marino and
G. Moore, \emph{"Some exact results on
tachyon condensation in string field theory,"}
\hepth{0009148}.
\bibitem{SenBT} D. Ghoshal and
A. Sen, \emph{"Normalisation of the Background 
Independent Open String Field Theory Action,"}
\hepth{0009191}.
\bibitem{Cornalba} L. Cornalba,
\emph{"Tachyon Condensation in Large Magnetic
Fields with Background Independent String Field
Theory,"} \hepth{0010021}.
\bibitem{Okuyama} K. Okuyama,
\emph{"Noncommutative Tachyon from Background
Independent Open String Field Theory,"} \hepth{0010028}.
\bibitem{MooreBT2} D. Kutasov, 
M. Marino and G. Moore,
\emph{"Remarks on Tachyon Condensation in
Superstring Field Theory,"} \hepth{0010108}.
\bibitem{Aleksejev} O. Andrejev,
\emph{"Some Computations of Partition Functions
and Tachyon Potentials in Background Independent
Off-Shell String Theory,"} \hepth{0010218}.
\bibitem{DasguptaBT1} S. Dasgupta  and T. Dasgupta,
\emph{"Renormalisation Group Analysis of Tachyon
Condensation,"} \hepth{0010247}.
\bibitem{Moriyama} S. Moriyama and S. Nakamura,
\emph{"Descent relation of tachyon condensation in
open boundary string theory,"} \hepth{0011002}.
\bibitem{ShatasviliBT3}
A. A. Gerasimov and S. L. Shatashvili,
\emph{"String Higgsy Mechanism and the Tate
of Open Strings,"} \hepth{0011009}.
\bibitem{Kleban} M. Kleban, A. Lawrence and S. Shenker,
\emph{"Closed strings from nothing,"} \hepth{0012081}.
\bibitem{KrausBST} P. Kraus and F. Larsen,
\emph{"Boundary String Field  Theory of $D\bar{D}$ system,"}
\hepth{0012198}.
\bibitem{TakayBT} T. Takayanagi, S. Terashima and
T. Uesugi, \emph{"Brane-Antibrane Action from
Boundary String Field Theory,"} \hepth{0012210}.
\bibitem{Andrejev} O. D. Andreev and A. A. Tseytlin,
\emph{"Partition Function Representation For The Open
Superstring Effective Action: Cancellation of Mobius 
Infinities And Derivative Corrections To Born-Infeld
Lagrangian,"} \npb{311}{1998}{205}.
\bibitem{Yasnov} D. Nemeschanski and V. Yasnov,
\emph{"Background Independent Open-String Field
Theory and Constant $B$-Field,"} \hepth{0011108}.
\bibitem{TseytlinSM} A. Tseytlin,
\emph{"Sigma model approach to string theory effective actions
for tachyons,"} \hepth{0011033}.
\bibitem{HarveyWS1} J. A. Harvey, S. Kachru, G. Moore and
E. Silverstein, \emph{"Tension is dimension,'}
\jhep{0003}{2000}{001}, \hepth{9909072}.
\bibitem{HarveyWS2} J. A. Harvey, D. Kutasov and E. Martinec,
\emph{"On the relevance of tachyons,"} \hepth{0003101}.
\bibitem{SenR} A. Sen, \emph{"Non-BPS States and branes in
string theory,"} \hepth{9904207}.
\bibitem{SenC} A. Sen, \emph{"Descent Relations among bosonic
D-branes,"} \ijmpa{14}{1999}{4061}.
\bibitem{WittenSFT} E. Witten,
\emph{"Noncommutative geometry and string field theory,"}
\npb{268}{1986}{253}.
\bibitem{SenFT1} A. Sen, 
\emph{"Universality of the tachyon potential,"}
\jhep{9912}{1999}{027}, \hepth{9911116}.
\bibitem{SenFT2} A. Sen and B. Zwiebach,
\emph{"Tachyon Condensation in String Field
Theory,"}
\jhep{0003}{002}{2000}, \hepth{9912249}.
\bibitem{BerkovitzFT1} N. Berkovits,\emph{"The Tachyon Potential in Open Neveu-
Schwarz String Field Theory,"} 
\jhep{0004}{022}{2000}, \hepth{0001084}.
\bibitem{HarveyFT} J. A. Harvey and P. Kraus,
\emph{"D-Branes as Lumps in Bosonic Open
String Field Theory,"}
\jhep{0004}{012}{2000}, \hepth{0002117}.
\bibitem{SenFT3} N. Berkovits, A. Sen and B. Zwiebach,
\emph{"Tachyon Condensation in Superstring Field
Theory,"} \hepth{0002211}.
\bibitem{TaylorFT} N. Moeller and W. Taylor,
\emph{"Level truncation and the tachyon in open
bosonic string field theory,"}
\npb{563}{105}{2000}, \hepth{0002237}.
\bibitem{KochFT} R. de Mello Koch, A.
Jevicki, M. Mihailescu and R. Tatar,
\emph{"Lumps and p-branes in open string field theory,"}
\plb{482}{249}{2000}, \hepth{0003031}.
\bibitem{Desmet} P. J. De Smet and 
J. Raeymaekers,
\emph{"Level-four approximation to the tachyon
potential in superstring field theory,"}
\jhep{0005}{051}{2000}, \hepth{0003220}.
\bibitem{NaqviFT} A. Iqbal and A. Naqvi,
\emph{"Tachyon Condensation On A Non-BPS
D-Brane,"} \hepth{0004015}.
\bibitem{SenFT4} N. Moeller, A. Sen and
B. Zwiebach, \emph{"D-branes as Tachyon Lumps
in String Field Theory,"} 
\jhep{0008}{039}{2000}, \hepth{0005036}.
\bibitem{DavidFT} J. R. David,
\emph{"$U(1)$ gauge invariance from open 
string field theory,"} \hepth{0005085}.
\bibitem{WittenFT} E. Witten,
\emph{"Noncommutative Tachyons And String Field
Theory,"} \hepth{0006071}.
\bibitem{RasteliFT} L. Rasteli and B. Zwiebach,
\emph{"Tachyon Potentials, Star Products and Universality,"}
\hepth{0006240}.
\bibitem{SenFT5} A. Sen and B. Zwiebach,
\emph{"Large Marginal Deformations in String
Field Theory,"} \hepth{0007153}.
\bibitem{TaylorFT2} W. Taylor,
\emph{"Mass generation from tachyon condensation
for vector fields on D-brane,"} \hepth{0008033}.
\bibitem{KochFT2} R. de Mello Koch and
J. P. Rodrigues, \emph{"Lumps in level truncated
open string field theory,"} \hepth{0008053}.
\bibitem{NaqviFT2} A. Iqbal and A. Naqvi,
\emph{"An Marginal Deformations in Superstring
Field Theory,"} \hepth{0008127}.
\bibitem{Kostelecky} A. Kostelecky and R. Potting,
\emph{"Analytical construction of a nonperturbative
vacuum for the open bosonic string,"}
\hepth{0008252}.
\bibitem{Schnabl} M. Schnabl,
\emph{"String field theory at large B-field 
and noncommutative geometry,"} \hepth{0010034}.
\bibitem{Rastelli} L. Rastelli, A. Sen and B. Zwiebach,
\emph{"String Field Theory Around the Tachyon
Vacuum,"} \hepth{0012251}.
\bibitem{Seiberg} N. Seiberg, \emph{"A Note on
Background Independence in Noncommutative Gauge Theories,
Matrix Model and Tachyon Condensation,"}
\jhep{0009}{2000}{003}, \hepth{0008013}.
\bibitem{Kluson1} J. Kluso\v{n}, 
\emph{"D-branes from $N$ Non-BPS D0-branes,"}
\jhep{0011}{2000}{016}, \hepth{0009189}.
\bibitem{Kluson2} J. Kluso\v{n},
\emph{"Matrix model and string field theory,"}
\hepth{0011029}. 
\bibitem{Terashima} S. Terashima,
\emph{"A Construction of Commutative D-branes 
from Lower Dimensional Non-BPS D-branes,"}
\hepth{0101087}.
\bibitem{TseytlinBI} A. A. Tseytlin,
\emph{"On non-abelian generalisation of Born-Infeld
action in string theory,"} \hepth{9701125}.
\bibitem{TaylorM1} W. Taylor,
\emph{"Lectures on D-branes, gauge theory and
M(atrices),"} \hepth{9801182}.
\bibitem{TaylorM2} W. Taylor,
\emph{"The M(atrix) model of M-theory,"}
\hepth{0002016}.
\bibitem{Schwarz} A. Konechny and A. Schwarz,
\emph{"Introduction to M(atrix) theory and noncommutative
geometry,"} \hepth{0012145}.
\bibitem{TaylorM3} W. Taylor,
\emph{"M(atrix) Theory: Matrix Quantum Mechanics as
a Fundamental Theory,"} \hepth{0101126}.
\bibitem{Ishibashi1} N. Ishibashi,
\emph{"p-branes from (p-2)-branes in the Bosonic 
String Theory,"} \npb{539}{1999}{107}, \hepth{9804163}.
\bibitem{Ishibashi2} N. Ishibashi,
\emph{"A Relation between Commutative and Noncommutative
Description of D-branes,"} \hepth{9909176}.
\bibitem{OkuyamaBS} K. Okuyama,
\emph{"Boundary States in B-Field Background,"}
\hepth{0009215}.
\bibitem{Alwis} S. P. de Alwis, \emph{"Boundary String Field Theory,
the Boundary State Formalism and D-Brane Tension,"}
\hepth{0101200}.
\bibitem{TseytlinDBI} A. A. Tseytlin,
\emph{"Born-Infeld action, supersymmetry and string
theory,"} \hepth{9908105}.
\end{thebibliography}
\end{document}